\documentclass[twocolumn, 11pt]{article}
\usepackage{array, float}
\usepackage[left=18mm, right=18mm, top=26mm, bottom=24mm]{geometry}
\usepackage{newtxtext}
\usepackage{amsmath, amssymb}
\usepackage{mathrsfs}
\usepackage{tabularx}
\usepackage{graphicx}
\usepackage{dcolumn}
\usepackage{setspace}
\usepackage{pgfpages}
\makeatletter
\usepackage{dcolumn}
\usepackage{url}
\usepackage{comment}
\usepackage{lineno}
\usepackage{cite}
\usepackage[numbers,sort&compress]{natbib}
\allowdisplaybreaks

\usepackage{abstract}

\usepackage{setspace}
\usepackage[font=small, format=plain, labelfont=bf, up, textfont=normal, up, justification=justified,singlelinecheck=false]{caption}

\usepackage[affil-it]{authblk}

\let\OLDthebibliography\thebibliography
\renewcommand\thebibliography[1]{
\OLDthebibliography{#1}
\setlength{\parskip}{0pt}
\setlength{\itemsep}{0pt plus 0.3ex}
}

\title{Evolution of cooperation in a three-strategy game combining snowdrift and stag hunt games}
\author{Hirofumi Takesue\thanks{Electronic address: \texttt{hir.takesue@gmail.com}}}
\affil{Tokyo Metropolitan University}
\date{}

\begin{document}

\twocolumn[

\maketitle

\begin{onecolabstract}
This study aimed to investigate the evolutionary dynamics of a three-strategy game that combines snowdrift and stag hunt games. This game is motivated by an experimental study, which found that individual solution lowers cooperation levels. Agents adopting this option aim to address a problem to the extent necessary to remove negative impact on themselves, although they do not free ride on cooperation effort provided by others. This property of the individual solution is similar to that of option defection in the stag hunt. Thus, the role of the interplay of defection in the snowdrift game and individual solution was examined in this study. The well-mixed population has two asymptotically stable rest points, one wherein the individual solution occupies the population, and the other wherein cooperation and defection coexist. The interactions on a square lattice enlarge the parameter region wherein cooperation survives, and the three strategies often coexist. The scrutinization of the evolutionary process shows that multiple mechanisms lead to the coexistence of the three strategies depending on parameter values. Our analysis suggests that considering the individual solution adds complexity to the evolutionary process, which might contribute to our understanding on the evolution of cooperation.
\\\\
\end{onecolabstract}
]
\saythanks

\section*{Introduction}
Research conducted over more than half a century has revealed that various mechanisms support the evolution of cooperation \citep{Nowak2006}. The prisoner's dilemma game is considered as one of the most canonical models of the problem of cooperation. The payoffs for mutual cooperation ($R$) are larger than those of mutual defection ($P$). However, unilateral defection achieves further individually beneficial outcomes ($T$) compared with mutual cooperation. In addition, the outcome of mutual defection is preferable to unilateral cooperation ($S$). The relationship $T > R > P > S$ characterizes the prisoner's dilemma and represents the contradiction between the social benefits of mutual cooperation and the temptation of free riding. Other games are also widely adopted to examine different social dilemmas, such as snowdrift (SD) and stag hunt (SH) games. The payoff order of the SD game is $T > R > S > P$, whereas that of SH game is $R > T \geq P > S$. These games have different issues compared with the prisoner's dilemma \citep{Doebeli2005, Skyrms2004}. Various mechanisms have been proposed, including kin selection \citep{West2007}, direct reciprocity in repeated interactions \citep{Hilbe2018b}, reputation \citep{Takacs2021, Xia2023}, selection of interaction partners \citep{Roberts2021}, punishment on non-cooperation \citep{Raihani2019}, and interactions on networks \citep{Szabo2007, Perc2017}, to explain the cooperation observed in societies.

Despite the progress in understanding the evolution of cooperation, recent experiments have suggested that the problem becomes complicated, and cooperation is impeded by the introduction of the third behavioral option, that is, individual solutions \citep{Gross2019, Gross2020}. People who adopt this third option do not free ride on the cooperation investment provided by others, but they try to solve the problem to the extent necessary to remove negative impact on themselves and do not contribute to collective solutions. Considering climate change, some people may try to collectively solve the problem itself, but others who adopt individual solutions may move away from susceptible areas and avert the damage of environmental problems \citep{Gross2020}. In their innovative studies, Gross and colleagues introduce experiments with three behavioral options \citep{Gross2019, Gross2020}. Investment into public pool corresponds to cooperation in classic social dilemmas, and all the members can enjoy the benefits provided by collective solutions if sufficient investment is made. Keeping resource corresponds to defection, and participants can enjoy the benefits of collective solutions for free (as long as the collective benefit is provided by others). The individual solution is a novel option. Each participant can ensure payoffs to a certain extent even the provision of collective solution has failed. This solution is less efficient than the successful provision of collective solutions because of the lack of scale merit. However, many participants in the experiment selected this safe option, and socially efficient outcomes achieved by the collective solution are impeded.

In this paper, a three-strategy game that combines the SD game and the SH game was considered to examine the role of the individual solution. We call this game the SDSH game. The original game introduced in the experiment has complex components such as repeated interactions and threshold of collective solutions. This study aims to examine the mechanism by which an individual solution affects the evolutionary dynamics by considering a stylized game. Two action options of the SDSH game correspond to investing in collective solutions and keeping resources in the experimental game. These two strategies are cooperation and defection in the classical SD game. Temptation to free riding urges to keep one's own resources, but failure to provide collective goods leads to serious ramifications. The individual solution is the third action, and its function is similar to defection in the SH game. The payoff for selecting the individual solution is smaller than that of mutual cooperation; thus, the individual solution is socially less efficient. However, the individual solution ensures constant payoffs regardless of the action of the partner, which prevents the solitary contribution to the collective solution. Therefore, cooperation and individual solution are similar to the two strategies in the SH game.

Analysis of the SDSH game shows two stable equilibria in a well-mixed population. In the first equilibrium, all the agents select the individual solution, and this equilibrium is always asymptotically stable. The second equilibrium is the mixture of cooperation and defection, and its stability depends on parameter values. Games in a structured population support cooperation, and all the strategies coexist in some stationary states. In addition, the coexistence of the three strategies is achieved by multiple mechanisms depending on parameter values. This analysis documents the role of a spatial structure in overcoming the reliance on the individual solution.

\section*{Model}
The SDSH game, which is a three-strategy game, has the properties of the SD and SH games. This game is represented by the following payoff matrix:
\[
\begin{array}{c}
C\\ D\\ I\\
\end{array}
\begin{pmatrix}
b-c & b-2c & b-2c \\
b   & 0 & 0 \\
b-c_I & b-c_I & b-c_I \\
\end{pmatrix},
\]
where $C$, $D$, and $I$ represent the cooperation, defection, and individual solution, respectively. Cooperation produces collective benefits, and both agents gain the benefit ($b$). When both agents select cooperation, the cost of cooperation is $c$. For simplicity, providing collective benefits costs two times as much as that in the case of unilateral cooperation. Agents who select defection enjoy the benefit without the cost of cooperation when the partner selects $C$. However, $D$ results in the worst payoff, 0, when the partner does not cooperate. Selecting the individual solution ensures a constant payoff: $b-c_I (> 0)$. The partner's cooperation does not have an effect because selecting $I$ already generates the required outcomes. We assume that $0 < c < c_I < 2c$, which indicates that the individual solution is costly than mutual cooperation, but solitary provision of the collective benefits costs even more. The SD game is recovered if we focus $C$ and $D$ under the assumption that $b > 2c$; the SH game is recovered if we focus $C$ and $I$. We suppose that $0 < c < c_I < 2c < b$, thereby summing up the assumptions introduced in this paragraph.

Evolutionary dynamics of the SDSH game is examined in well-mixed and structured populations \citep{Szabo2007, Perc2017}. In the well-mixed population, the evolutionary process follows replicator dynamics. The rest points and their stability are examined. The frequencies of the three strategies in equilibrium states ($\rho_C$, $\rho_D$, and $\rho_I$) are the primarily reported outcomes. In the structured population, the evolution proceeds on a square lattice with periodic boundary conditions. The following Monte Carlo simulations are conducted to examine the evolutionary dynamics. In each elementary timestep, one agent, $i$, is randomly selected out of $N = L^2$ agents; one of the $i$'s neighbor, $j$, is also randomly selected. The agent $i$ may imitate the strategy of the agent $j$ based on payoff comparison. Both agents participate in the games with their four neighbors and accumulate payoffs ($\Pi_i$ and $\Pi_j$). The probability of strategy imitation is calculated as $1 / [1 + \exp(\beta (\Pi_i - \Pi_j))]$, where $\beta$ refers to the intensity of selection. Mutations occur with a small probability of $\mu$, and the agent $i$ adopts a randomly selected strategy. In a Mote Carlo step (MCS), each agent has one opportunity to update the strategy on average. Agents’ strategies are assigned randomly at the initial state. Strategy frequencies in stationary states are also denoted as $\rho_C$, $\rho_D$, and $\rho_I$. In typical simulations, the length of the relaxation period is set to $2 \times 10^5$ MCSs and that of the sampling period is set to $2 \times 10^4$ MCSs. The mean values of four simulation runs are reported to enhance statistical accuracy. In verifying the outcomes of the well-mixed population, a similar simulation is conducted on a complete network.

\section*{Results}
\begin{figure}[tbp]
\centering
\vspace{5mm}
\includegraphics[width = 85mm, trim= 0 0 0 0]{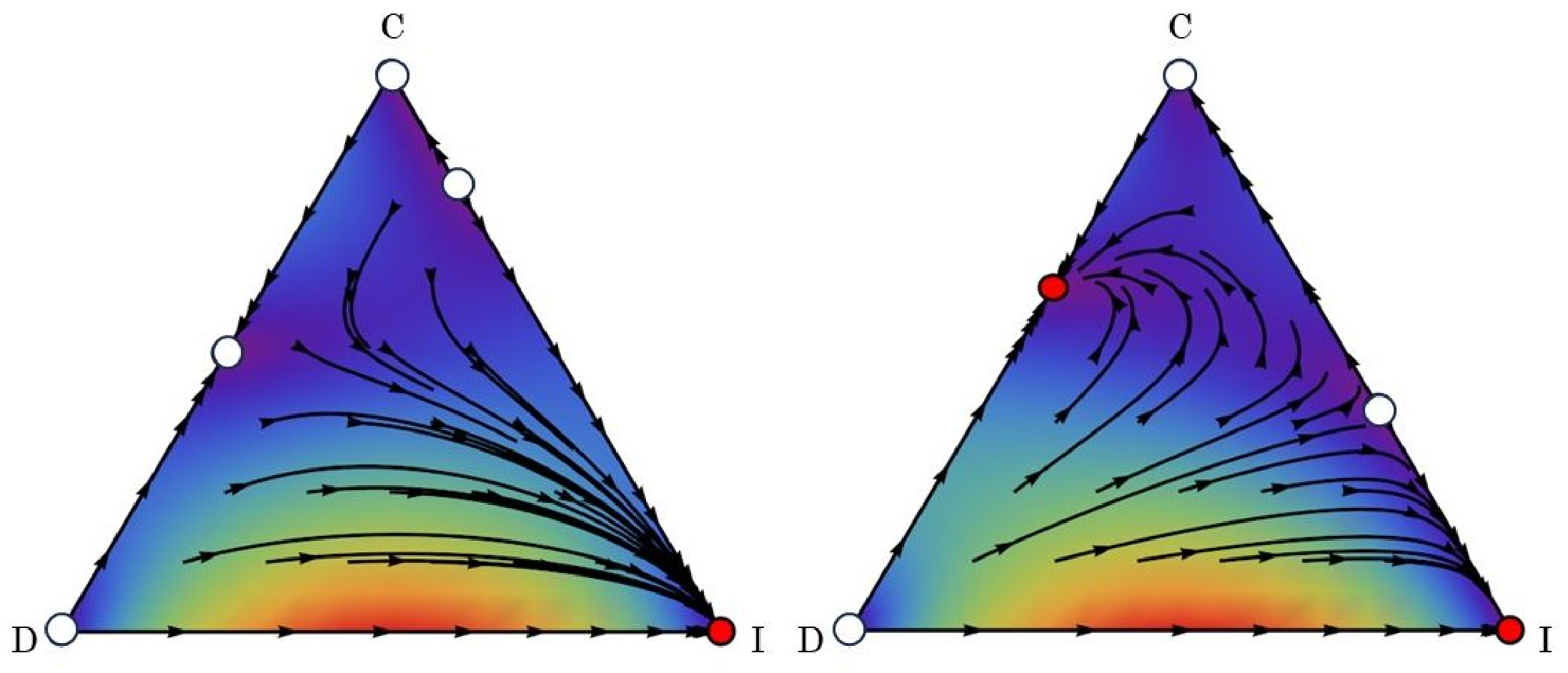}
\caption{\small Replicator dynamics of the SDSH game. The left panel shows that the asymptotically stable rest point is dominated by the individual solution ($b = 3, c = 1$, and $c_I = 1.2$). The right panel shows that the mixed strategy equilibrium consisting of cooperation and defection is also stable ($b = 3.6, c = 1$, and $c_I = 1.6$). This figure is produced by the software used in the visualization of evolutionary dynamics \citep{Izquierdo2018}.}
\label{SDSH_wellmixed}
\end{figure}

Replicator dynamics has two asymptotically stable equilibria. In the first equilibrium, all the agents adopt the individual solution. Stability is confirmed by examining the payoff matrix; selecting $I$ achieves the largest payoffs when interacting with the partner who selects $I$. This equilibrium is stable regardless of the parameter values (Figure~\ref{SDSH_wellmixed}). The second equilibrium consists of cooperation and defection, in which the cooperation frequency is $(b - 2c) / (b - c)$. The stability of this equilibrium depends on parameter values; this equilibrium is stable in the right panel but not in the left panel.

\begin{figure}[tbp]
\centering
\vspace{5mm}
\includegraphics[width = 85mm, trim= 0 0 0 0]{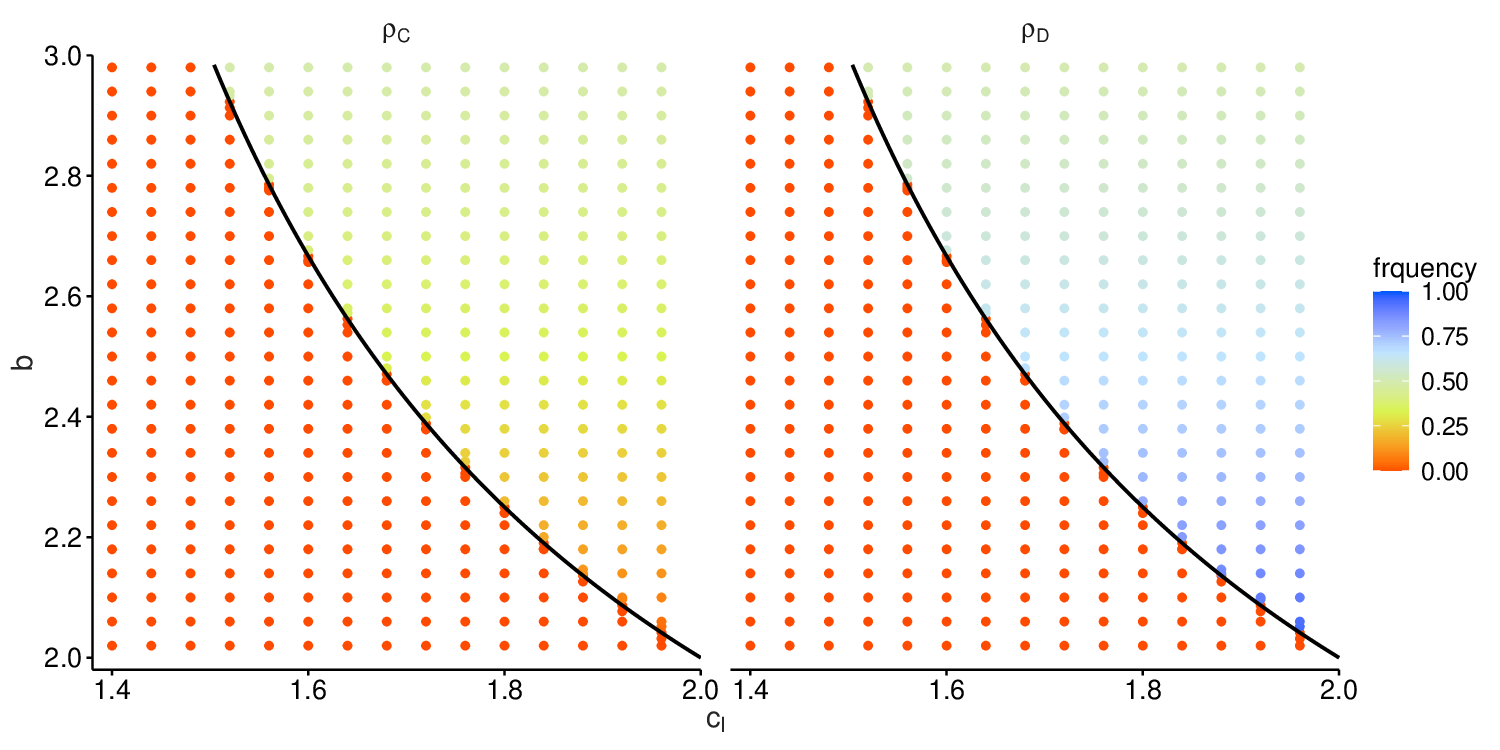}
\caption{\small Strategy frequencies observed in the agent-based simulations on the complete network. The curve represents the relationship $(b - c)(c_I - c) = c^2$. The lower left parameter region of the curve corresponds to a region wherein the mixed strategy equilibrium is unstable, whereas the upper right region corresponds to a region wherein it is asymptotically stable. Parameters: $N = 120^2, c = 1, \beta = 10$, and $\mu = 10^{-4}$.}
\label{phase_rho_sigma_b}
\end{figure}

The stability of the second equilibrium can be depicted by using a simple condition. In examining the equilibrium stability, agent-based simulations were conducted on a complete network where the initial condition is this equilibrium. Figure~\ref{phase_rho_sigma_b} shows the resultant strategy frequencies of cooperation and defection. When the equilibrium is unstable, the population evolves to the state wherein (almost) all the agents adopt the individual solution (the lower left region of the curve). By contrast, the equilibrium strategy frequencies are maintained with some parameter values. The calculation shown in the Appendix indicates that this equilibrium is asymptotically stable when $(b - c)(c_I - c) > c^2$, which is visualized by the upper right region of the curve in the figure.

\begin{figure}[tbp]
\centering
\vspace{5mm}
\includegraphics[width = 85mm, trim= 0 0 0 0]{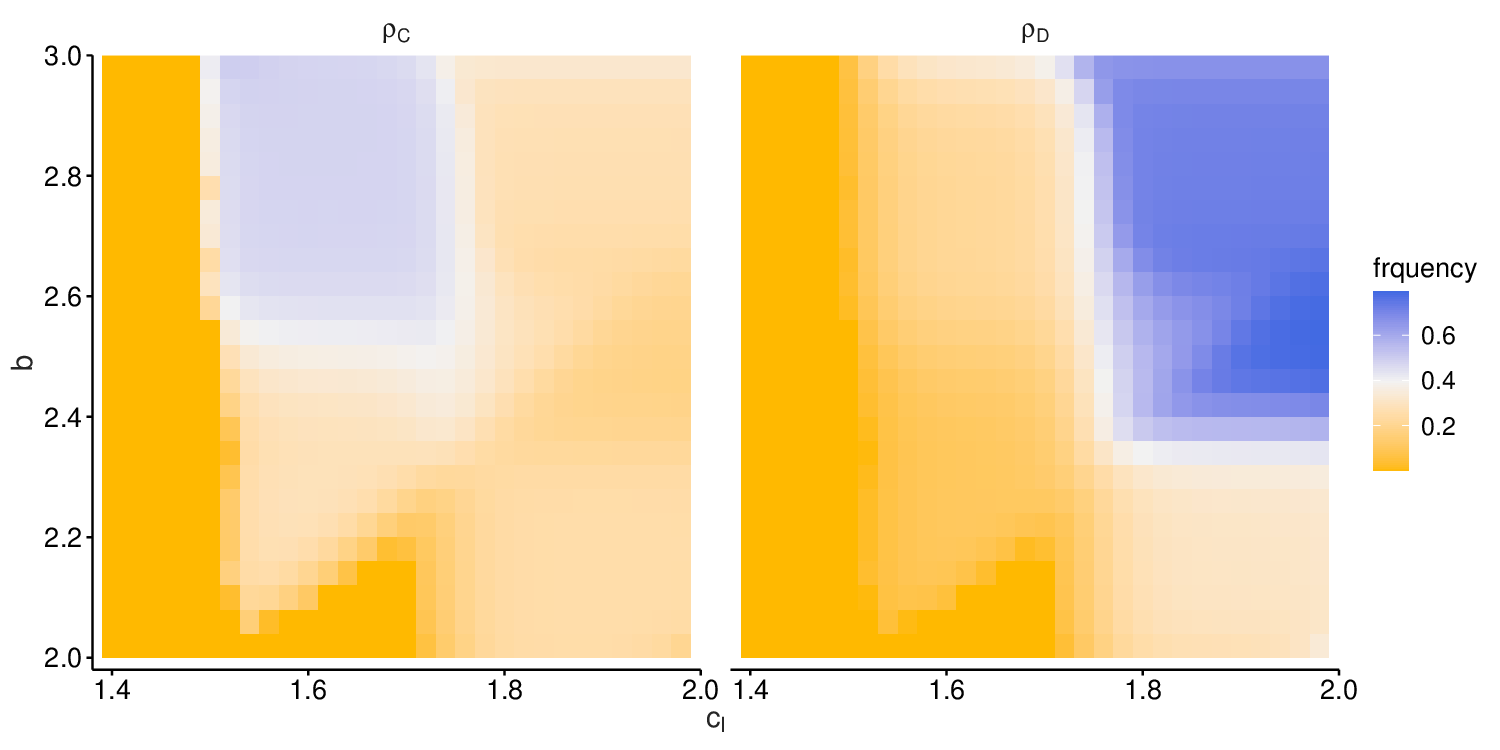}
\caption{\small Frequencies of cooperation and defection as a function of the benefit of cooperation ($b$) and the cost of the individual solution ($c_I$). $\rho_C$ reaches the maximum values with moderate values of $c_I$ when $b$ is large. Otherwise, the system shows more complex behavior; the coexistence of the three strategies is observed with moderate and large values of $c_I$. Parameters: $L = 120, c = 1, \beta = 10$, and $\mu = 10^{-4}$.}
\label{phase_rho_c_b}
\end{figure}

The interaction on a square lattice shows more complex outcomes, including the coexistence of the three strategies (Figure~\ref{phase_rho_c_b}). The small cost of the individual solution leads to the dominance of the individual solution as in the well-mixed population; both values ($\rho_C$ and $\rho_D$) are negligible in the left regions of the panels. The increase in the cost of the individual solution leads to the stationary state wherein all the three strategies coexist. The qualitative patterns of the frequencies of the three strategies depend on the values of $b$. When the benefit is large ($b \geq 2.32$ in the figure), $\rho_D$ increases monotonically with the value of $c_I$, whereas $\rho_C$ reaches the maximum value with moderate values of $c_I$. The strategy frequencies show a more complex pattern with a small benefit ($2.06 \leq b \leq 2.3$ in the figure). The value of $\rho_C$ reaches the local maximum, but the further increase in $c_I$ leads to the re-emergence of the increase in the frequencies of the individual solution. (Both $\rho_C$ and $\rho_D$ are negligible.) In some cases ($b \leq 2.16$), the individual solution almost occupies the population again. However, the coexistence of the three strategies is observed again with the further increase of $c_I$. Notably, the coexistence of the three strategies can be ascribed to the lattice, as no corresponding equilibrium is observed in the well-mixed population.

\begin{figure}[tbp]
\centering
\vspace{5mm}
\includegraphics[width = 85mm, trim= 0 0 0 0]{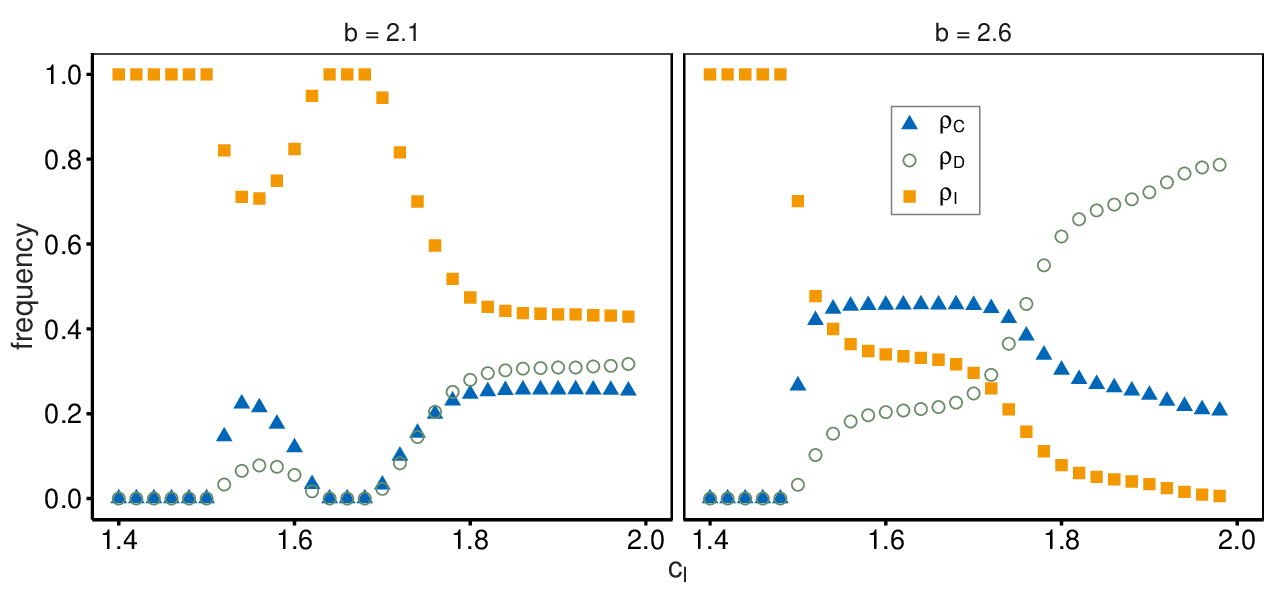}
\caption{\small Frequencies of the three strategies as a function of $c_I$ for different values of $b$. The patterns observed in Figure~\ref{phase_rho_c_b} are replicated using a larger system size. For $b = 2.1$, the system size is 600 when $c_I = $ 1.52, 1.62, and 1.7. For $b = 2.6$, the system size is 480 when $c_I = 1.5$. A larger system size (e.g., $L = 800$) reaches the same results. In other cases, $L = 120$. Parameters: $c = 1, \beta = 10$, and $\mu = 10^{-4}$.}
\label{rho_c_b}
\end{figure}

As shown in Figure~\ref{rho_c_b}, the same patterns are observed when using a larger system size (up to $L = 600$). When $b = 2.6$, $\rho_C$ reaches the maximum value with intermediate values of $c_I$, whereas the values of $\rho_D$ ($\rho_I$) monotonically increase (decrease). By contrast, when $b = 2.1$, the values of $\rho_C$ and $\rho_D$ show a local maximum. The frequencies of cooperation and defection are sizable compared with the mutation probability ($\mu = 10^{-4}$). In addition, the dominance of the individual solution and the coexistence of the three strategies are observed two times each in different parameter regions. Our next analysis focuses on the mechanism of the rise and fall of the strategy frequencies.

\begin{figure*}[tbp]
\centering
\vspace{5mm}
\includegraphics[width = 120mm, trim= 0 0 0 0]{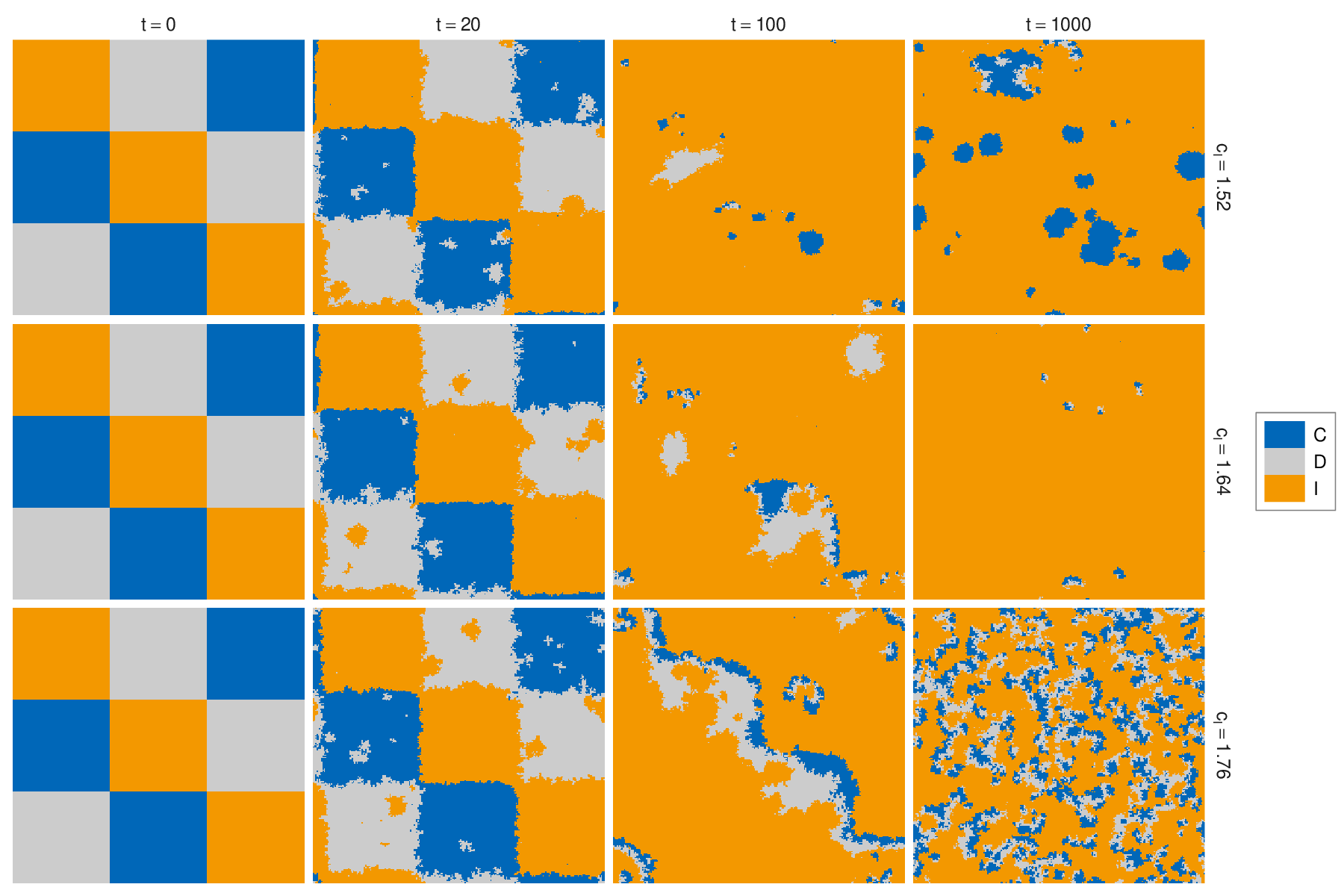}
\caption{\small Evolutionary processes on a square lattice for different values of the cost of the individual solution. The suppression of defection by the individual solution supports the survival of cooperative clusters ($c_I = 1.52$). This mechanism does not work when the cost of the individual solution is high because the remaining agents adopting defection will destroy cooperation clusters ($c_I = 1.64$). Further increase in the cost of individual solution manifests the advantage of cooperation over the individual solution, which leads to typical cyclic dominance. Parameters: $L = 240, b = 2.1, c = 1, \beta = 10$, and $\mu = 10^{-4}$.}
\label{lattice_time}
\end{figure*}

Visual inspection of the lattice shows that the coexistence of the three strategies is achieved with different spatial dynamics. Figure~\ref{lattice_time} shows the distribution of the three strategies at different costs of the individual solution when $b = 2.1$. Simulation starts with the specified conditions to elucidate the spatial patterns ($t = 0$). Both the cases for $c_I = 1.52$ and $c_I = 1.64$ show the proliferation of the individual solution when $t = 100$, but the spatial arrangement of the remaining cooperation and defection shows different patterns. In the case of $c_I = 1.52$, defection is sufficiently suppressed, and some clusters of cooperators do not accompany neighbors adopting defection. By contrast, when the cost of the individual solution increases to 1.64, the insufficient suppression and mutation allow the survival of defection, and the remaining cooperators accompany neighbors adopting defection. This difference becomes evident when $t = 1000$. The remaining cooperators maintain some clusters when $c_I = 1.52$. Conversely, cooperation and defection decrease their frequencies when $c_I = 1.64$. The remaining agents that adopt defection continuously utilize cooperators, but defection vanishes because the exploited cooperators do not survive. In the case of $c_I = 1.76$, the advantage of cooperation over the individual solution becomes more evident, and cooperation survives despite defection. Consequently, cyclic dominance is observed wherein agents employing $C$ adopt $D$; agents employing $D$ adopt $I$, and agents employing $I$ adopt $C$. This cyclic dominance can be confirmed by the upper rightward wave mainly observed in the center of the figure when $t = 100$. For $c_I = 1.52$ and $c_I = 1.76$, similar cooperation levels are observed, but the spatial configuration implies different underlying mechanisms.

\begin{figure}[tbp]
\centering
\vspace{5mm}
\includegraphics[width = 85mm, trim= 0 0 0 0]{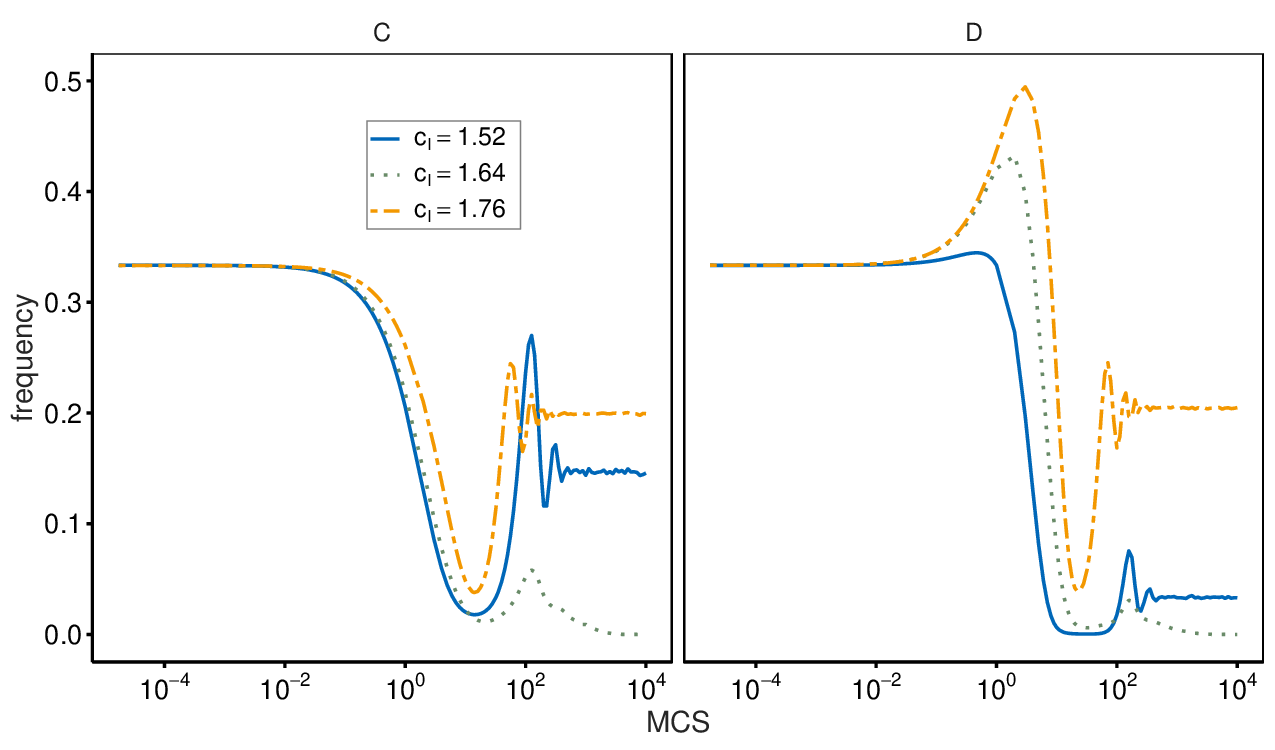}
\caption{\small Time-series changes in strategy frequencies. At $c_I = 1.52$, defection almost disappears, which leads to increased cooperation and defection. This pattern is contrary to that at $c_I = 1.64$, where the dominance of the individual solution is observed. At $c_I = 1.76$, larger frequencies of cooperation and defection are observed, which corresponds to cyclic dominance. The reported values are the average values of 1 000 simulation runs. Parameters: $L = 240, b = 2.1, c = 1, \beta = 10$, and $\mu = 10^{-4}$.}
\label{SDSH_time}
\end{figure}

Time-series changes in strategy frequencies also corroborate this pattern (Figure~\ref{SDSH_time}). For $c_I = 1.52$ and $c_I = 1.64$, similar patterns are observed, but differences are observed in the degree of decrease in defection frequencies observed after its initial increase. Agents adopting defection almost vanish when $c_I = 1.52$, but small frequencies of defection remain when $c_I = 1.64$. This pattern is consistent with the pattern shown in Figure~\ref{lattice_time}; the small clusters of cooperators accompany neighboring agents who adopt defection at $c_I = 1.64$ but not at $c_I = 1.52$. This difference confirms whether or not cooperation (and defection) frequencies later turn into a significant increase. Cooperation evolves when defection has been sufficiently suppressed. The case for $c_I = 1.76$ shows larger cooperation frequencies than the other cases. Cyclic dominance ensures cooperation over the individual solution, which leads to relatively large frequencies of cooperation and defection.

\section*{Discussion}
In this study, the evolutionary dynamics of a three-strategy game that combines the SD and the SH games is examined, which is motivated by experimental studies that demonstrate the critical role of the individual solution in social dilemma \citep{Gross2019, Gross2020}. Our study shows two asymptotically stable equilibria in a well-mixed population under the replicator dynamics; the first equilibrium is dominated by the individual solution, and the second equilibrium is the mixed-strategy equilibrium of cooperation and defection. The interaction on a networked population enlarges the parameter region wherein cooperation survives, and the coexistence of the three strategies emerges. The spread of cooperation enjoys the benefits of restricted interactions on networks. In addition, the visualization of the system indicates that multiple mechanisms work in sustaining the coexistence of all strategies depending on parameter values.

The SDSH game introduced in this study is akin to the game with an option of voluntary participation \citep{Szabo2002a, Wu2005, Guo2017, Wang2018a, Hu2020, You2020, Zhang2020a, Gao2021a}. In particular, the loner strategy that permits the voluntary participation is similar to the individual solution in this study. Although many studies focus on voluntary prisoner's dilemma games, a study has examined the voluntary SD game \citep{Zhong2007}. The SDSH game and voluntary SD game show similar behavior in a well-mixed population. There are two asymptotically stable equilibria; the first one is the mixed equilibrium of cooperation and defection, and the second one is dominated by the third strategy (the loner strategy or the individual solution). By contrast, a structured population shows a qualitatively different pattern. In particular, the dominance of the individual solution and the coexistence of the three strategies are observed two times each in different parameter regions, which is unique to the SDSH game. By contrast, the effect of the loner's payoff in the voluntary SD game is almost monotonic, and the cooperation levels show simpler behavior. The unreported simulations of the voluntary SD game with mutation replicate this pattern. The coexistence of three strategies has two types, which are related to the complex pattern in the SDSH game. In the first type, sufficient suppression of defection leads to coexistence. In the second type, typical cyclic dominance supports the coexistence, which is also observed in the voluntary social dilemma \citep{Szabo2002a, Cardinot2019, Hu2020, You2020, Gao2021a, Jia2024}. The local maximum of cooperation levels and cyclic dominance are also observed in prisoner's dilemma game with an exit option \citep{Shen2021}. The game is equivalent to the combination of the prisoner's dilemma and the SH game if the terminology of this paper is applied (i.e., the PDSH game). Despite the different game structure, similar dynamics emerges. The motivation of this study is also akin to the framework of multi-games \citep{Wang2014, Deng2018, Guo2019, Li2019a, Deng2021b, Li2021b}. In typical multi-games, each agent participates in two different games, and payoffs depend on the game type they participate in. The heterogeneity introduced by the game types often enhances the cooperation. Considering multiple games by considering a third strategy can also enrich the understanding on the evolutionary dynamics in social dilemma.

The limitations of this study are also discussed. First, spatial interaction is only one of the solutions to the social dilemma. The role of different mechanisms such as repeated interactions \citep{Yamamoto2019}, reputation \citep{Podder2021a}, mobility \citep{Cardinot2019}, a different updating rule \citep{Lu2017}, and the introduction of the fourth strategy \citep{Guo2020c} is considered in voluntary social dilemma. In addition, the motivating study extends the original experiment by considering the role of punishment \citep{Gross2019}, which is an ongoing research topic in theoretical literature \citep{Chen2015, Szolnoki2017a, Yang2018, Wang2019d, Chowdhury2021, Flores2021, Sun2021, Wang2021h, Lee2022a, Zhou2022, Hua2023a, Lv2023, Quan2023, Wang2023a, Zhu2023a, Lee2024}. Whether or not different mechanisms support cooperation in the SDSH game should be elucidated. Second, this study does not consider group interaction. A simple two-person game is utilized in this study, but group interaction shows unique behavior that cannot be reduced to a simple game \citep{Perc2013a}. The public goods game may be a natural extension of this study. Finally, our study assumed a square lattice, but literature has documented the role of other lattices and heterogeneous networks \citep{Santos2005, Roca2009, Flores2022}. Square lattices are the canonical setting \citep{Perc2017}, but the careful consideration of different network structures might be necessary. These extensions will capture the properties of actual social dilemmas in the society as well as the complicated games in the experiment more accurately.

\section*{Appendix}
The behavior of a well-mixed population is discussed in this section. First, we explain the condition that the mixed strategy equilibrium of cooperation and defection is asymptotically stable. In this section, the dynamically evolving proportion of agents who employ cooperation, defection, and individual solution is denoted as $x$, $y$, and $z$, respectively. As $x = 1 - y - z$, the dynamics of $y$ and $z$ is examined. Payoff values for each strategy are calculated as follows:
\begin{eqnarray*}
\pi_C &=& b - c - (y + z)c  \\
\pi_D &=& b(1 - y - z)    \\
\pi_I &=& b - c_I.
\end{eqnarray*}
Using these values, the following replicator equations are considered:
\begin{eqnarray*}
\dot{y} &=& y(b(1 - y - z) - \bar{\pi})    \\
\dot{z} &=& z(b - c_I - \bar{\pi}),
\end{eqnarray*}
where $\bar{\pi} = x \pi_C + y \pi_D + z \pi_I$. At the equilibrium ($y^* = c/(b-c), z^* = 0$), the elements of the Jacobian matrix are given as follows:
\begin{eqnarray*}
J_{11}^* = \left. \frac{\partial \dot{y}}{\partial y} \right|_{y = y^*, z = z^*} &=& 
-\frac{c(b - 2c)}{b-c}    \\
J_{12}^* = \left. \frac{\partial \dot{y}}{\partial z} \right|_{y = y^*, z = z^*} &=&
\frac{c}{b - c} \left(\sigma - \frac{b^2 - 2bc + 2c^2}{b - c}\right)    \\
J_{21}^* = \left. \frac{\partial \dot{z}}{\partial y} \right|_{y = y^*, z = z^*} &=&
0    \\
J_{22}^* = \left. \frac{\partial \dot{z}}{\partial z} \right|_{y = y^*, z = z^*} &=&
-\frac{bc_I - bc - c_Ic}{b - c}. 
\end{eqnarray*}
The sign of the two eigenvalues, $J_{11}^*$ and $ J_{22}^*$, determine the stability of this equilibrium. The value of $J_{11}^*$ is negative because of the assumption on the values of payoff-related parameters, which indicates that the stability of the equilibrium can be determined by the sign of $J_{22}^*$. Consideration of the assumption that $b - c > 0$ leads to the inequality presented in the Results section.

Second, the inner equilibria, that is, the equilibria wherein all the three strategies coexist, are briefly discussed. When the inner equilibria exist, the components of the equilibria wherein the strategy frequencies are $x^{**} = (b - 2c)/(b-c)$ and $y^{**} + z^{**} = 1 - x^{**}$ are observed. For the payoffs of the three strategies to be equal and the equilibria exist, the following equality must hold: $bc_I - bc - c_Ic = 0$. The inner equilibria require the equality that includes all the payoff-related parameters to hold. Given this limited condition, the inner equilibria are empirically irrelevant in the SDSH game. Thus, we focus on the other cases and equilibria.


\end{document}